\documentclass[aps,pre,twocolumn,showpacs,groupedaddress]{revtex4}  
\usepackage{graphicx}  
\usepackage{dcolumn}   
\usepackage{bm}        
\usepackage{amssymb}   
\usepackage{float}
\usepackage{multirow}
\usepackage{enumerate}

\begin{document}
\hspace{5.2in}

\title{Chimera-like states in two distinct groups of identical populations of coupled Stuart-Landau oscillators}
\author{K. Premalatha$^{1}$, V. K. Chandrasekar$^{2}$, M. Senthilvelan$^{1}$, M. Lakshmanan$^{1}$}
\address{$^1$Centre for Nonlinear Dynamics, School of Physics, Bharathidasan University, Tiruchirappalli - 620 024, Tamil Nadu, India.\\
$^2$Centre for Nonlinear Science \& Engineering, School of Electrical \& Electronics Engineering, SASTRA University, Thanjavur -613 401,Tamilnadu, India.}
\begin{abstract}
We show the existence of chimera-like states in two distinct groups of identical populations of globally coupled Stuart-Landau oscillators.  The existence of chimera-like states occurs only for a small range of frequency difference between the two populations and these states disappear for an increase of mismatch between the frequencies.  Here the chimera-like states are characterized by the synchronized oscillations in one population and desynchronized oscillations in another population.  We also find that such states observed in two distinct groups of identical populations of nonlocally coupled oscillators are different from the above case in which coexisting domains of synchronized and desynchronized oscillations are observed in one population and the second population exhibits synchronized oscillations for spatially prepared initial conditions.  Perturbation from such spatially prepared initial condition leads to the existence of imperfectly synchronized states.  An imperfectly synchronized state represents the existence of solitary oscillators which escape from the synchronized group in population-I and synchronized oscillations in population-II.  Also the existence of chimera state is independent of the increase of frequency mismatch between the populations.  We also find the coexistence of different dynamical states with respect to different initial conditions which causes multistability in the globally coupled system.  In the case of nonlocal coupling, the system does not show multistability except in the cluster state region. 
\end{abstract}

\pacs{05.45.Xt, 89.75.-k}
\maketitle
\section{Introduction} 
\par Dynamical systems in nature are rarely isolated.  Interactions between the dynamical units often give rise to new phenomena which can be exploited in different contexts in physical, biological and social sciences \cite{1}.  Among them, the recently discovered emergent phenomenon of chimera states has been in the center of recent research towards the study of coupled networks.  In the chimera state the synchronous and asynchronous behaviors are observed simultaneously in a network of coupled identical oscillators.  Chimera state was first reported by Kuramoto et al. in nonlocally coupled phase oscillators \cite{2,3}.  The fascinating phenomenon was initially found in nonlocal coupling configurations \cite{14,15,23,24,25,26,27,28,29,30,31,32}.  However more recent results reveal that the systems with globally \cite{4,4a,4b,4c} and locally coupled oscillators \cite{37ll} are also capable of showing such phenomenon.   Recently such states have also been explored in experiments with optical \cite{33}, electronic \cite{8}, electrochemical \cite{6}, and mechanical oscillators \cite{7}.
\par  Such a remarkable phenomenon has been subsequently studied for a variety of systems including chemical oscillators \cite{34a}, neuron models \cite{28n}, planar oscillators\cite{37}, heterogeneous systems \cite{36}, oscillators with more than one populations \cite{38b,38cc,38cr,ana,37t1,38b1,37ph} as well as in hierarchical networks \cite{37Hi} and also in a variety of system structures such as rings, networks with two and three oscillator populations \cite{11,r1} and two dimensional map lattices \cite{25}.  Several other chimera patterns like chimera death states \cite{38c, 4b, 38t, 38cp} and phase-flip chimera states \cite{38pf} have been recently explored.  For instance, in Ref. \cite{14}, Abrams et al. studied the stability of the chimera states in two interacting populations of globally coupled phase oscillators where one population is synchronized and the other population is desynchronized.  It does not contain synchronized oscillators from both the populations.  They have also shown that the chimera states are stable and stationary, and that these states become periodically breathing chimera states due to stability change.  Later Pikovsky and Rosenblum \cite{38b} have observed quasiperiodic chimera states in the same model considered in Ref. \cite{14} by nonuniformly distributing the initial conditions.  On the other hand, globally clustered chimera (GCC) states were reported by Sheeba, Chandrasekar and Lakshmanan  \cite{26} in the case of delay coupled phase oscillators. They have shown that the existence of such states can be achieved by tuning the delay parameter. 
\par Whereas many of the studies related to the chimera patterns have been focusing on the networks of phase oscillators, in this work we address the question as to how the emergence of chimeras arise in two distinct groups of identical populations of Stuart-Landau oscillators where variations in amplitudes also occur.   Here the interaction between the oscillators are equally shared by both the populations.  One important fact to be noted here is that the frequencies of the oscillators are identical within their population but have a finite difference among the two groups.  Our system is different from the system considered in Ref. \cite{14} where the authors assumed that the oscillators within a population are coupled strongly with each other than that with the neighboring population and also differ from the system considered in Ref.\cite{36}, where Laing assumed the two populations of phase oscillators with nonidentical natural frequencies.  Recently, in ref. \cite{r7}, the authors have identified transition from cluster states to extensive chaos and then to incoherent state while decreasing the system parameter in an identical population of globally coupled Stuart-Landau oscillators.  
\par Our motivation in this paper is to investigate the occurrence of chimera-like states by considering two distinct groups (each one is characterized by its own frequency) of identical populations of Stuart-Landau oscillators which are coupled through (i) global and (ii) nonlocal couplings.  We study the characteristic behavior of the oscillators by fixing the frequency of the oscillators in one population while varying the frequency of the oscillators in the other population.  We find the occurrence of different dynamical states including local synchronization, local cluster states, chimera-like states, global cluster states and global synchronization in globally coupled oscillators.  Among these dynamical states, chimera-like states occur only for a sufficient range of frequency mismatch and all the oscillators are coupled through equal coupling interaction.  In contrast to the chimera states found in a single array of coupled identical oscillators, the present case of chimera-like states (where synchronized oscillations occur in one population while another population is in a fully desynchronized state) exist in two distinct groups of identical populations with a frequency mismatch between them.  It occurs neither for a very low mismatch nor for a larger mismatch of frequencies.  These states are different from chimera states found in coupled identical oscillators in which total number of oscillators are split into coexisting domains of synchronized and desynchronized oscillations \cite{4b}. 
\par Further, we also find that the chimera states exhibiting synchronized and desynchronized oscillations in nonlocally coupled two distinct groups of identical populations are different from the globally coupled case for the reason that in the previous case, synchronized and desynchronized oscillators exist in one population while another population shows synchronized oscillations.  Existence of such a state is independent of frequency mismatch between the two populations under nonlocal coupling and exist for spatially prepared initial conditions.  Perturbation from such initial states leads to imperfectly synchronized states in this region.  In globally coupled system, we find the multistability in the system depending on the initial conditions used.  On the other hand, a nonlocally coupled system does not show any multistability region except in the cluster region.
\begin{figure*}[ht!]
\begin{center}
\includegraphics[width=1.0\linewidth]{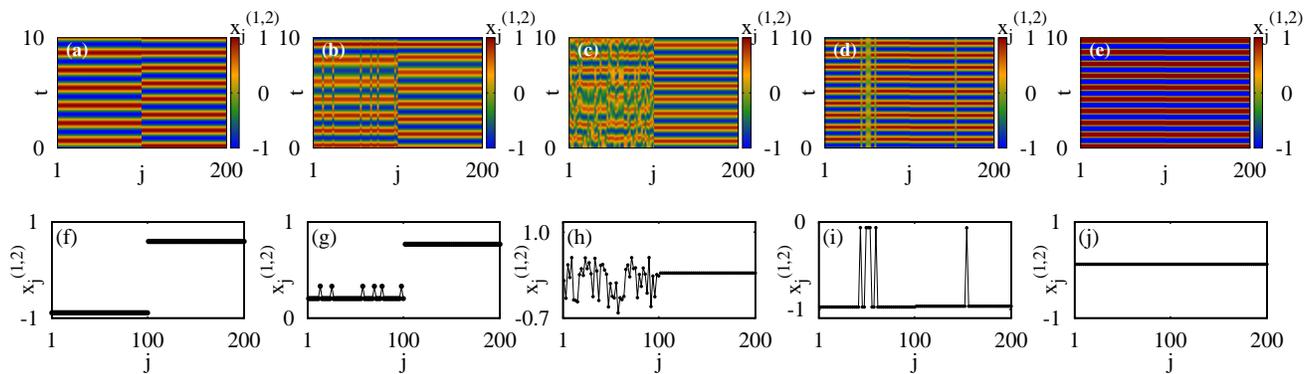}
\end{center}
\caption{(Color online) Space-time plots of the variables $x_j^{(1,2)}$ depicting (a) local synchronization for $\varepsilon=0.1$, (b) local cluster synchronization for $\varepsilon=0.8$, (c) chimera-like state for $\varepsilon=1.1$, (d) global cluster synchronization for $\varepsilon=1.8$, (e) global synchronization for $\varepsilon=2.8$.  In all these cases we have chosen $\Delta \omega=0.5$ and $c=3.0$.  Further, (f)-(j) represent the corresponding snapshots of the variables $x_j$ for Figs. (a)-(e).  Note that in the above, $x_j^{(1,2)}$, j=1,2,...,100 correspond to population-I, that is $x_j^{(1)}$, and $x_j^{(1,2)}$, j=101,102,...,200 correspond to population-II, that is $x_j^{(2)}$.  One can obtain similar plots for $y_j ^{(1,2)}$ also, which are not presented here.}
\label{global}
\end{figure*} 
\begin{figure}[ht!]
\begin{center}
\includegraphics[width=1.0\linewidth]{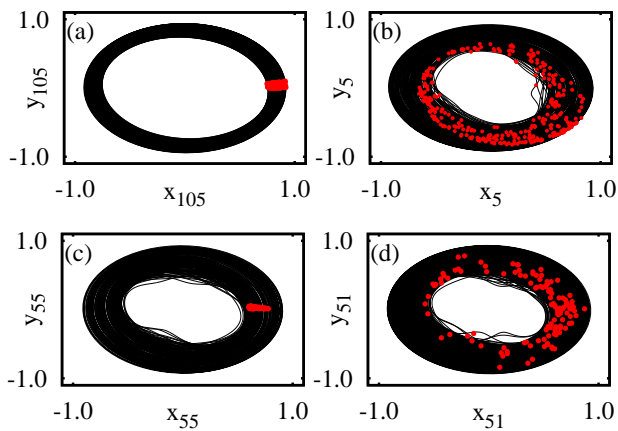}
\end{center}
\caption{(Color online) Phase portraits of the representative oscillators in chimera-like states: (a) For synchronized oscillator ($z_{105}$) with segment $\Lambda_{105}$ (red/grey dots), (b) for desynchronized oscillator ($x_{5}$) with data set $A_5$ (red/grey dots), (c) for desynchronized oscillator ($x_{55}$) with segment $\Lambda_{55}$ (red/grey dots), (b) for desynchronized oscillator ($z_{51}$) with data set $A_{51}$ (red/grey dots).}
\label{phasep}
\end{figure} 
\par The structure of the paper is organized as follows:   In Sec. II we describe our model of two distinct groups of identical populations of globally coupled oscillators.  In Sec. III we discuss the existence of various dynamical states by introducing the frequency mismatch between the populations in the above globally coupled oscillators with two parameter phase diagrams.  In Sec. IV we discuss how the dynamics of nonlocally coupled oscillators get affected by the presence of frequency mismatch between them.  Finally we critically summarize our results in Sec. V.
\section{MODEL OF Two DISTINCT competing identical populations of Stuart-Landau oscillatorS}
\par We consider two distinct populations of globally coupled Stuart-Landau oscillators for our analysis,  by assuming the same common frequencies within a population and different frequencies between the populations.  The governing equations are specified by the following set of equations,
 \begin{eqnarray}
\dot{z_j}^{(1,2)}=(1+i\omega^{(1,2)})z_{j}^{(1,2)}-(1- ic)|z_{j}^{(1,2)}|^2 z_{j}^{(1,2)}\nonumber\\+\varepsilon (\overline{z}-z_{j}^{(1,2)}),\qquad \qquad\qquad
\label{glo}
\end{eqnarray}
where $z_{j}^{(1,2)}=x_{j}^{(1,2)}+iy_{j}^{(1,2)}$, $\overline{z}=(1/N)\sum_{k=1}^{N}z_k$. There are $M_1$ oscillators in the population-I and they are numbered as $z_j^{(1)}$ while $M_2$ oscillators are in the population-II and they are numbered as $z_j^{(2)}$.  $M_1+M_2=N$  and $N$ is the total number of oscillators.  In our study, we consider both the populations to be of equal size, $M_1=M_2=M$.  For a discussion on $M_1\ne M_2$, see the last paragraph of Sec. III below.  In Eq. (\ref{glo}), $\varepsilon$ is the coupling constant and $c$ is the nonisochronicity parameter, $\omega^{(1)}$ and $\omega^{(2)}$ are the natural frequencies of the oscillators in the two populations of the network.  This model can also be considered as a single population with two clustered networks.  In our simulations, we choose the number of oscillators $N$ to be equal to 200 and in order to solve the equation (\ref{glo}), we use the fourth order Runge-Kutta method with time step 0.01 and the initial state of the oscillators ($x_j^{(1,2)},~y_j^{(1,2)}$) are distributed with uniform random values between -1 and +1.  We have also verified that the results are independent of increasing the number of oscillators in the population. 
\section{Global interaction} 
\par To start with, by assuming $\omega_1\ne 0$, $\omega_2\ne 0$ in (\ref{glo}), we consider a small frequency difference between the two populations such that $\omega^{(2)}-\omega^{(1)}=\Delta\omega=0.5$.  To explore the results of different dynamical behaviors, we present the associated space-time plots in Figs. \ref{global}(a-e).  For small values of the coupling strength of the oscillators, both the populations are individually synchronized with two different average frequencies (frequency profiles of the oscillators are discussed below in Fig. \ref{freq}), see Figs. \ref{global}(a) and \ref{global}(f).  This means that the oscillators are having the same average frequency within their population but they are different for different populations.   By increasing the coupling strength to $\varepsilon=0.8$ the system of oscillators in population-I splits into two groups.  These two groups are oscillating synchronously with same frequency though there is a finite phase difference between the two groups, while the population-II remains oscillating synchronously.  They are clearly illustrated with the space-time plot in Fig. \ref{global}(b) and with a snapshot of the variables in Fig. \ref{global}(g).  Hence these states are designated as local cluster states.  Upon increasing the strength of coupling interaction to $\varepsilon=1.1$ one observes that the mismatch between the frequencies of two populations causes asynchronous oscillations among the oscillators in population-I while the population-II remains synchronized which denotes the existence of chimera-like states.  In contrast to the chimera states found in a single array of coupled identical oscillators, the present case of chimera-like states exist in two identical populations with a frequency mismatch between them.  Figure \ref{global}(c) represents the space-time plot for the chimera-like state.  Its snapshot in Fig. \ref{global}(h) confirms the coexistence of spatially coherent and incoherent distributions of oscillators.  In this state the high frequency population remains synchronized having the same average frequency while the low frequency population is desychronized with its oscillators having different average frequencies. 
 On increasing $\varepsilon$ to $\varepsilon=1.8$ we note that the system enters into the global cluster state as shown in Fig. \ref{global}(d).  In this state the total number of oscillators in both the populations are grouped into two groups (Fig. \ref{global}(i)).  These two groups are entrained to a common average frequency but with different amplitudes.  We can observe the globally synchronized oscillations (with same frequency, amplitude and phase) among the oscillators for $\varepsilon$ beyond $\varepsilon=2.2$. 
\par On order to confirm the existence of chimera-like states, we analyze the phase-locking behaviour of the synchronized and desynchronized oscillators through the following localized set approach \cite{local}.  To illustrate this approach, we choose one of the representative oscillators from both the synchronized and desynchronized groups.  First, we construct a data set $A_5$ by observing the desynchronized oscillator $z_5$ whenever the trajectory of synchronized oscillator $z_{105}$ crosses the segment $\Lambda_{105}~(={x_{105}>0.8,y_{105}\approx 0.0})$.  In Fig. \ref{phasep}(a), the attractor of synchronized oscillator $z_{105}$ is shown by black curve and red/grey line represents the segment $\Lambda_{105}$.  The attractor of the desynchronized oscillator $z_5$ is depicted by black curve and the data set $A_5$ is shown by the red/grey dots in Fig. \ref{phasep}(b) and in which the spreading of the data set over the trajectory clearly illustrates that the deysnchronized oscillator is not in phase with the synchronized oscillator.  Further, we also illustrate the non-phase locking behaviour between the desynchronized oscillators in Figs. \ref{phasep}(c) and (d).  The attractor and corresponding segment of one of the randomly chosen representative oscillators $z_{55}$ are shown by the black curve and red/grey line in Fig. \ref{phasep}(c).  In Fig. \ref{phasep}(d), we can observe that the data set $\Lambda_{51}$ corresponding to the oscillator $z_{51}$ is spread over the trajectory $z_{51}$.  This also clearly illustrates that there is no phase locking between the desynchronized oscillators.  Thus we confirm that the existence of chimera-like states which is illustrated in Figs.  \ref{global}(c) and (h).  The chimera-like states are distinguished from the intermittent chimera states \cite{r2} and quasi-periodic chimera states \cite{3} that are observed in two populations of coupled rotators where the authors have considered broken-symmetry conditions realized by initializing the first population with identical phases and frequencies while they are random for second population.  Note that the important fact to be noted here is that the frequencies of the oscillators are identical within their population but have a finite difference among the groups.  Also the initial conditions are chosen uniformly between $-1$ and $+1$.

\begin{figure}[ht!]
\begin{center}
\includegraphics[width=1.0\linewidth]{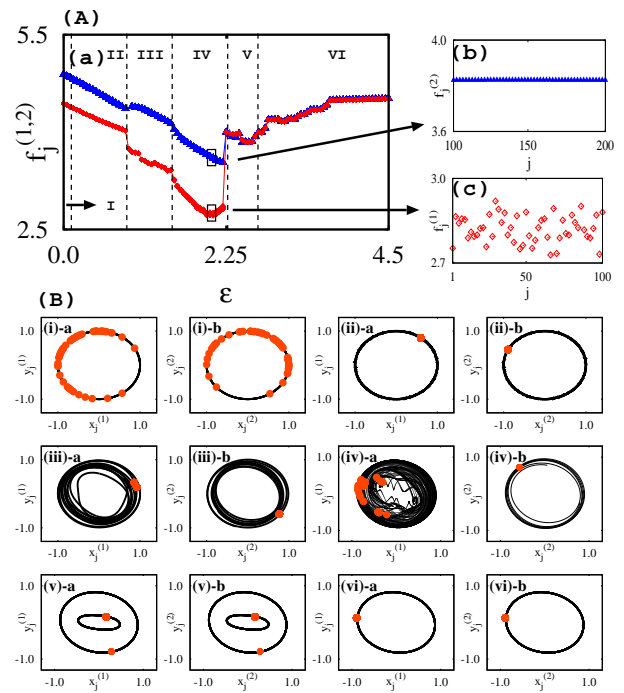}
\end{center}
\caption{(Color online) A(a) Average frequencies of the oscillators as a function of the coupling strength for the parameter values $\Delta \omega=0.5$, $c=3.0$.  Enlarged structures for a specific value of $\varepsilon=2.05$: (b) Average frequencies of the synchronized oscillators (population-II). (c) Frequencies of the desynchronized oscillators (population-I).  B. Corresponding to the evolution in the complex plane, snapshots of the variables $(x_j^{(1,2)},y_j^{(1,2)})$: (i)-a and b correspond to desynchronization for $\varepsilon=0.01$ (Region-I), (ii)-a, b correspond to local synchronization for  $\varepsilon=0.1$ (Region-II), (iii)-a, b represent local cluster states for $\varepsilon=0.8$  (Region-III), (iv)-a, b represent the chimera-like states for $\varepsilon=1.1$ (Region-IV), (v)-a, b represent the global cluster states for $\varepsilon=1.8$ (Region-V) (vi)-a, b corresponds to global synchronization fro $\varepsilon=2.8$ (Region-VI).  Trajectories are shown by black solid lines and red/grey dots represent the snapshots of the variables $z_{j}^{(1,2)}$ in the complex plane. }
\label{freq}
\end{figure} 
\begin{figure*}[ht!]
\begin{center}
\includegraphics[width=1.0\linewidth]{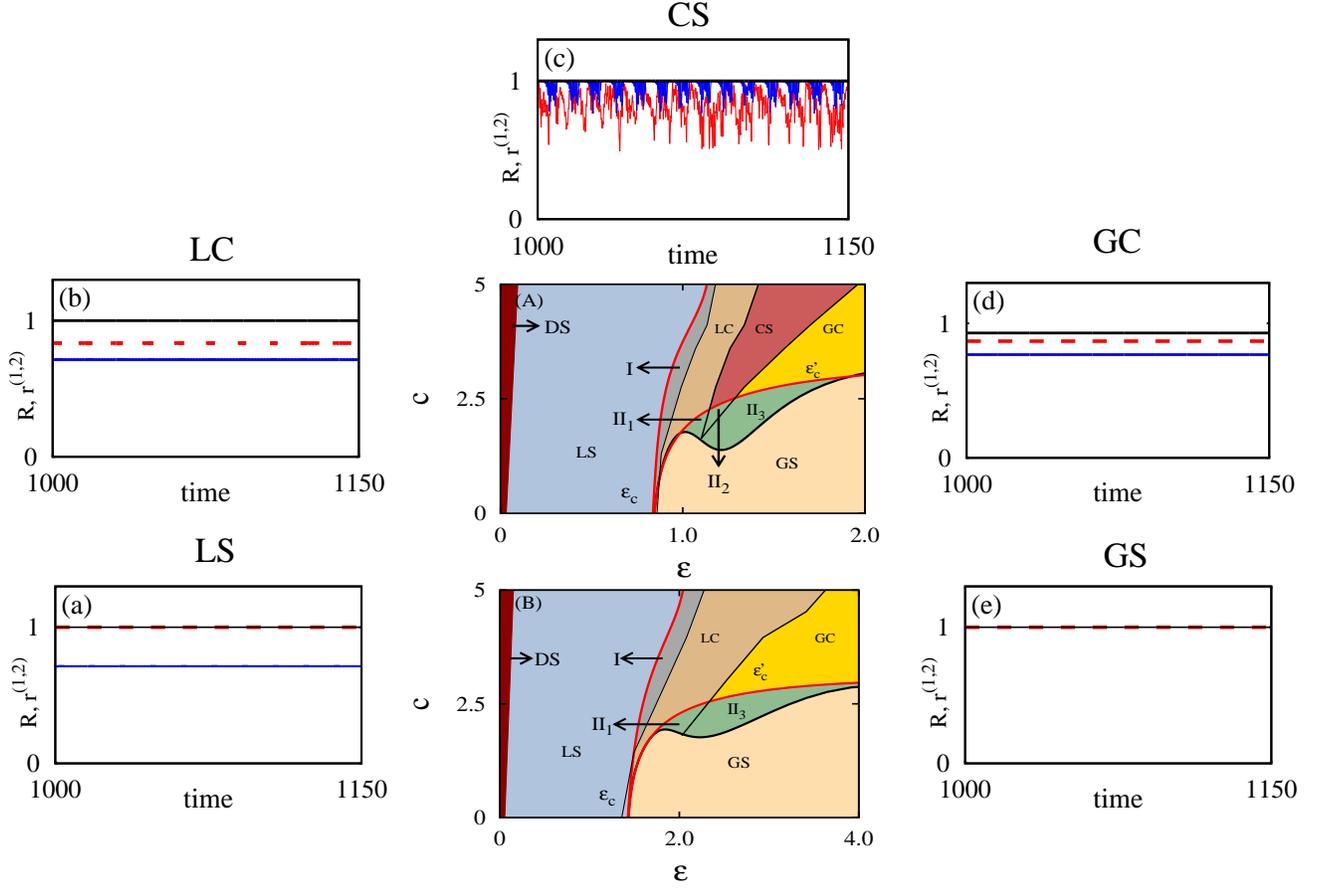}
\end{center}
\caption{(Color online) The two parameter phase diagram in the ($\varepsilon, c$) parametric space: (A) for $\Delta\omega=0.5$ and (B) for $\Delta\omega=1.5$.  Boundary $\varepsilon_c$ is obtained from the critical value of the coupling strength given in Eq. (\ref{cc}).  The curve $\varepsilon_c'$ is obtained from the Eq. (\ref{ev}) and the other boundaries are obtained numerically.  Region LS represents the local synchronization, region GS represents the global synchronization, region LC indicates the local cluster states, region CS represents the chimera-like states and region GC represents the global cluster states.   Region -I is the multistability region between the local synchronization and local cluster states.  Regions $II_1$ is the multistability region between complete synchronization and local cluster states.  Region $II_2$ is the multistability region between complete synchronization and chimera states.  Region $II_3$ is the multistability region between complete synchronization and global cluster states.  Figs. (a) - (e) present the global synchrony order parameter R (blue/dark-grey solid curve) and local order parameter $r^{(1)}$ (red dashed curve) and $r^{(2)}$ (black solid curve) for the local synchronization, local cluster states, chimera-like states, global cluster states and global synchronization, respectively.}
\label{a1}
\end{figure*}

\par To give more details about the nature of the dynamical states (discussed above), we present the frequency of the oscillators as a function of the coupling strength ($\varepsilon$) in Figs. \ref{freq} (A).  We calculate the average frequencies of the oscillators by using the expression $f_j=2 \pi \Omega_j/\Delta T $, where $\Omega_j$ is the number of maxima in the time series $x_j$ over a time interval $\Delta T$.  Also for a clear understanding about the dynamical regions, we have plotted the evolution of the dynamical variables in the complex plane $(z_j^{(1,2)}=x_j^{(1,2)}+iy_j^{(1,2)})$ and their respective snapshots of the variables $(x_{j}^{(1,2)},y_{j}^{(1,2)})$ are projected onto the complex plane in Fig. \ref{freq} B(i-vi).  In Fig. \ref{freq} A(a), initially for the range between $ 0.0<\varepsilon< 0.05$, the oscillators are having same frequency within populations but they are oscillating incoherently as shown in Figs. \ref{freq}B(i)-a and (i)-b.  For the range  $ 0.05<\varepsilon< 0.8$, we can observe local synchronizations, that is each population is individually synchronized with two different frequencies represented by region-II in Fig. \ref{freq} A(a).  Figs. \ref{freq} B(ii)-a and (ii)-b clearly show the individual synchronization.  In the region $0.8<\varepsilon<1.4$, we can observe local cluster states.  In this state, oscillators in the low frequency population separate into different synchronized groups (Figs. \ref{freq} B(iii)-a and (iii)-b).   But their frequencies are the same while the high frequency population remains synchronized with a single frequency as shown in the region-III in Fig. \ref{freq} A(a).  Upon increasing $\varepsilon$ to $1.4<\varepsilon<2.25$, we can observe chimera-like states in the region-IV in Fig. \ref{freq} A(a) where the coherent oscillators are oscillating with the same frequency while the incoherent oscillators are having different frequencies which are clearly shown in Figs. \ref{freq} A(b) and A(c).  Figs. \ref{freq} B(iv)-a and (iv)-b also validate the chimera-like behaviour where the incoherent oscillators are randomly distributed in the complex plane while the distribution of the synchronized oscillators are shown by the red/grey dot.  Further for the values of $\varepsilon$ between 2.25 to 2.8, one can find global cluster states where the frequencies of all the oscillators (in both the populations-I and II) are the same as illustrated in region-V  of Fig. \ref{freq} A(a) (and further confirmed by Figs. \ref{freq} B(v)-a and (v)-b).  Finally, beyond the value $\varepsilon= 2.8$, both the populations are globally synchronized to a common frequency, see region VI in Fig. \ref{freq}A(a) and Figs. \ref{freq} B(vi)-a and (vi)-b.

\par By considering the various dynamical states described above, except for the chimera-like state and global cluster state, all the oscillators in each of the populations possess the same amplitude (note that in the local cluster region also the amplitudes $r_j= \sqrt{x_j^2+y_j^2}$ remain the same).  Hence in order to study analytically the regions of local and global synchronizations, we assume that the amplitudes of all the oscillators remain the same to find the regions of local and global synchronizations.
By substituting $z_j^{(1,2)}=r_j^{(1,2)}e^{i\theta_j^{(1,2)}}$ in equation (1) we will get

 \begin{eqnarray}
\dot{r_j}^{(1,2)}=r_j^{(1,2)}-r_j^{3(1,2)}\qquad \qquad \qquad\qquad\qquad\qquad \nonumber\\+\frac{\varepsilon}{N}\sum_{k=1}^N(r_k^{(1,2)} \cos(\theta_k^{(1,2)}-\theta_j^{(1,2)})-r_j^{(1,2)})\nonumber\\
\dot{\theta_j}^{(1,2)}=\omega^{(1,2)}+c r_j^{2(1,2)}\qquad \qquad \qquad\qquad\qquad\qquad\nonumber\\+\frac{\varepsilon}{N}\sum_{k=1}^N \frac{r_k^{(1,2)}}{r_j^{(1,2)}}\sin(\theta_k^{(1,2)}-\theta_j^{(1,2)})
\label{rthe}
\end{eqnarray}
If we substitute $r_k^{(1,2)}=r_j^{(1,2)}=constant$ so that $\dot{r_j}^{(1,2)}=0$, we will get the phase equation as 
\begin{eqnarray}
\dot{\theta_j}^{(1,2)}=\omega^{(1,2)}+c(1-\varepsilon)+\frac{\varepsilon c}{N }\sum_{k=1}^N \cos(\theta_j^{(1,2)}-\theta_k^{(1,2)})\nonumber\\+\frac{\varepsilon }{N }\sum_{k=1}^N \sin(\theta_j^{(1,2)}-\theta_k^{(1,2)}).
\label{pe}
\end{eqnarray}
  Complex order parameter within each population is described by $z^{(1,2)}=r^{(1,2)}e^{i\psi{(1,2)}}=(1/M)\sum_{k=1}^Me^{i\theta_k^{(1,2)}}$.  Replacing the summation term by the complex order parameter, equation (\ref{pe}) becomes, 
\begin{eqnarray}
\dot{\theta_j}^{(1,2)}=\omega^{(1,2)}+c(1-\varepsilon)+\frac{\varepsilon c}{N }r^{(1,2)}\cos(\theta_j^{(1,2)}-\psi^{(1,2)})\nonumber\\+\frac{\varepsilon }{N }r^{(1,2)}\sin(\theta_j^{(1,2)}-\psi^{(1,2)}).
\label{rt}
\end{eqnarray}
 In the case of local synchronization region (LS), oscillators within each group are identical and so the local order parameter $r^{(1,2)}=1$ (red dashed and black solid line) while the global order parameter $R<1$ (blue/dark-grey solid line) which is shown in Fig. \ref{a1}(a).  Note that the global order parameter is defined through $Z=Re^{i\Psi}=\frac{1}{2}\sum_{\sigma=1}^2r^{(\sigma)}e^{i\psi{(\sigma)}}$, where $R$ measures the degree of global synchrony for the entire system.  In the region GS, we can observe the global synchronization and the values of both $R$ and $r^{(1,2)}$ are unity as shown in Fig. \ref{a1}(e).  The critical value of coupling strength at which global synchronization occurs is given by \cite {ana}
\begin{equation}
\varepsilon_c=\Delta\omega/(2 \cos\alpha),
\label{cc}
\end{equation}
 where $\alpha=arc\tan[c]$. From the linear stability analysis of Eq. (\ref{rt}), we will get the eigenvalues for in-phase synchronized states \cite {ana} as 
\begin{equation}
\lambda_{\pm}=-\varepsilon \cos \alpha-\varepsilon \cos(\pm\Delta\psi+\alpha) <0,
\label{ev}
\end{equation}    
where $\Delta\psi=arc\sin[\Delta\omega/(2\varepsilon \cos\alpha)]$.  The choice $\lambda_+=0$ leads to the curve $\varepsilon_c'$ shown by the red/grey curve in Fig. \ref{a1}(A) and it matches with the numerically obtained boundary.  Different dynamical regions in Fig. \ref{a1}(A) for $\Delta\omega=0.5$ and (B) for $\Delta\omega=1.5$ are identified with the help of strength of incoherence \cite{38} for every choice of $\varepsilon$ and $c$ values.  Some details on strength of incoherence are given in Appendix A.  Fig. \ref{a1}(a) represents local synchronization where the local order parameters $r^{(1)}=1$ (black solid curve), $r^{(2)}=1$ (red dashed curve) and global order parameter $R<1$ (blue/dark-grey solid curve). In the regions corresponding to local cluster states (LC), chimera-like states (CS) and global cluster states (GC), we can observe the local order parameters $r^{(2)}=1$ (black solid curve), $r^{(1)}<1$ (red dashed curve) and global order parameter $R<1$ (blue/dark-grey solid curve) which are depicted with Figs. \ref{a1}(b), (c) and (d). In globally synchronized region, both $r^{(1,2)}$ and $R$ take the value unity which is shown in Fig. \ref{a1}(e).  The region-I is the multistability region between local synchronization and local cluster states.  The region-$II_1$ is the multistability region between the local cluster states and complete synchronization.  Also there arises a multistability region between chimera-like states and complete synchronization in region-$II_2$.  Similarly, region-$II_3$ represents the multistability region between the global cluster states and complete synchronization.  The regions LS, LC, CS, GC and GS are always stable, that is in these regions even for a small perturbation of the initial state of the oscillators from the completely synchronized solution leads to the regions LS, LC, CS, GC and GS, respectively.  There also arises a question whether the chimera-like states are robust to increasing or decreasing the mismatch of frequencies.  To address this question, we now present the existence of dynamical states in the parametric space ($ \varepsilon,\Delta \omega$).
\begin{figure}[ht!]
\begin{center}
\includegraphics[width=1.0\linewidth]{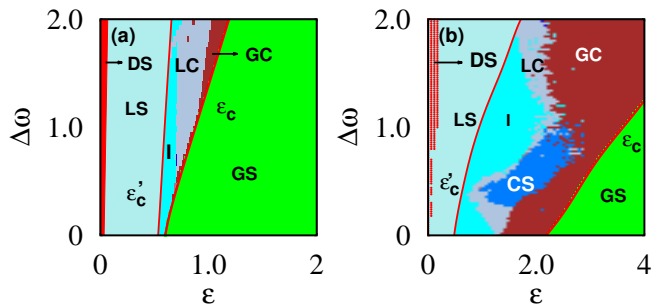}
\end{center}
\caption{(Color online) Phase diagrams for globally coupled system (\ref{glo}) by varying the values of $\Delta \omega$ and  $\varepsilon$ (a) for $c=1.5$ and (b) for $c=3.0$.  Boundary $\varepsilon_c$ is obtained from the critical value of the coupling strength given in Eq. (\ref{cc}).  The curve $\varepsilon_c'$ is obtained from Eq. (\ref{ev}) and the other boundaries are obtained numerically.  The region DS represents the desynchronized state, LS is the local synchronization region, GS shows the globally synchronized state, GC shows the state corresponding to global cluster region, CS shows chimera states, and LC shows the local cluster state.  The region-I is the multistability region between the local synchronization and local cluster states.}
\label{a2}
\end{figure} 
\begin{figure}[ht!]
\begin{center}
\includegraphics[width=1.0\linewidth]{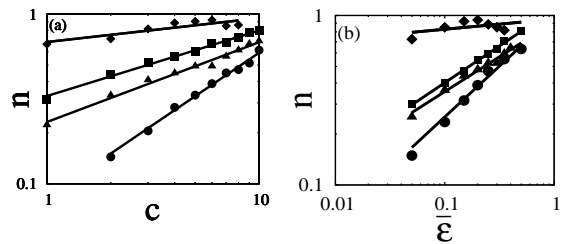}
\end{center}
\caption{Log-log plots: (a) The value of $n$ (= $\frac{M_1}{M_2}$) as a function of the nonisochronicity parameter $c$  for fixed  $\varepsilon$ values ($\varepsilon=0.7$ for local synchronization, $\varepsilon=0.9$ for local cluster states, $\varepsilon=1.5$ for chimera-like states, and $\varepsilon=2.3$ for global cluster states) and (b) the value of $n$ (= $\frac{M_1}{M_2}$) as a function of the coupling strength $\bar{\varepsilon}$ for a fixed $c=3.0$.  Note that $\bar{\varepsilon}=\varepsilon-\varepsilon_e$,  where $\varepsilon_e$ is the earliest coupling strength.  Dots ($\bullet$) represent the numerical data and the corresponding best fit is represented by the (black) curve for the chimera-like states.  Similarly the local synchronization, local cluster states and global cluster states are represented by the solid line with ($\blacktriangle$), ($ \blacklozenge$) and ($\blacksquare$), respectively.  Note that $\varepsilon_e=0.05$ for local synchronization, $\varepsilon_e=0.8$ for local cluster state, $\varepsilon_e=1.4$ for chimera-like state and $\varepsilon_e=2.25$ for global cluster state.  Other parameter value: $\Delta\omega=0.5$}
\label{log}
\end{figure} 
\begin{figure*}[ht!]
\begin{center}
\includegraphics[width=1.0\linewidth]{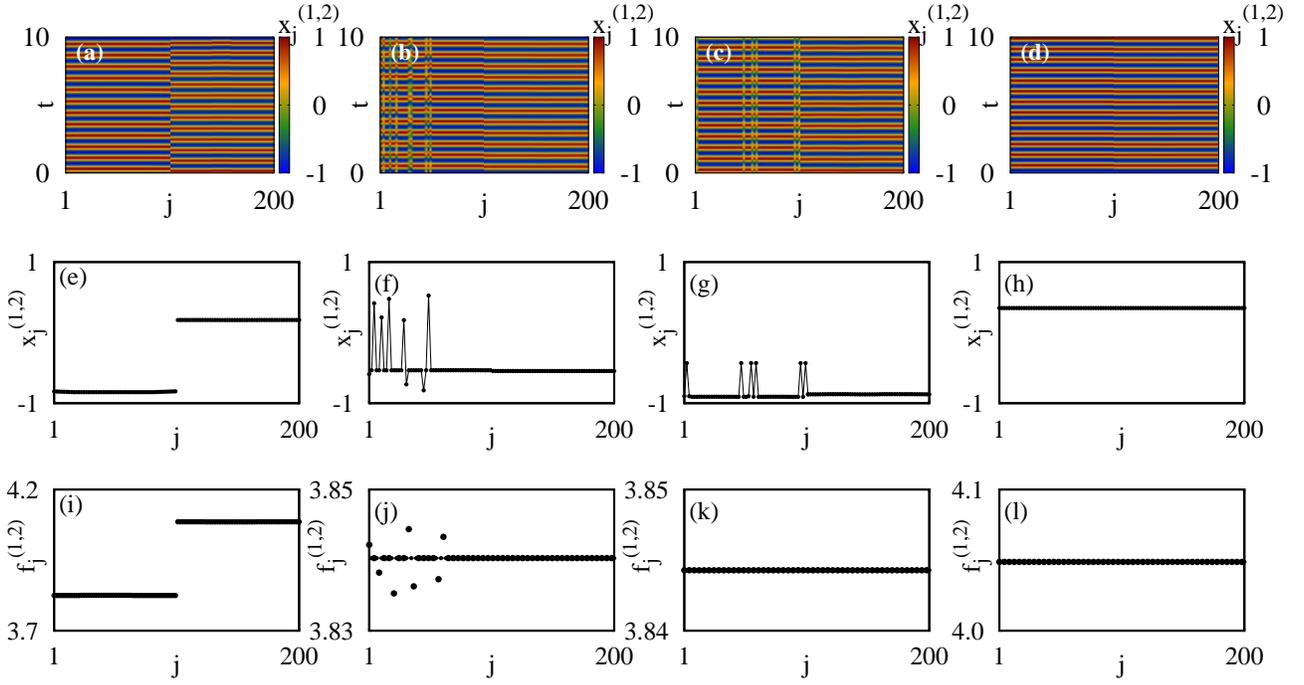}
\end{center}
\caption{(Color online) Space-time plots of the variables $x_j^{(1,2)}$: (a) local synchronization for $\varepsilon=0.2$, (b) imperfectly synchronized state (solitary state) for $\varepsilon=0.35$, (c) cluster state for $\varepsilon=0.4$, (d) global synchronization for $\varepsilon=1.1$ with $c=3.0$,$r=0.4$ and $\Delta \omega=0.5$.  Figs. (e)-(h) represent the snapshots of the variable $x_j^{(1,2)}$ for Figs. (a)-(d).  Figs.  (i)-(l) represent the corresponding frequency profiles of the oscillators for Figs. (a)-(d).  Again here oscillators j=1,2,...,100 correspond to group-I ($x_j^{(1)}$) while oscillators numbered j=101,102,...,200 represent group-II ($x_j^{(2)}$).}
\label{n1}
\end{figure*} 
\par In order to analyze the global picture of the dynamical system under the influence of frequency mismatch, we plotted the two parameter phase diagrams in the parametric space ($\varepsilon,\Delta\omega$) for two different nonisochronicity parameter values, that is $c=1.5$ and $c=3.0$, which are illustrated in Figs. \ref{a2}(a) and (b), respectively.  In these figures, the curve $\varepsilon_c$ is obtained from the critical value of the coupling strength for globally synchronized region given in Eq. (\ref{cc}).  The curve $\varepsilon'_c$ is obtained from Eq. (\ref{ev}) and the other boundaries are obtained numerically.  These regions are identified with the help of strength of incoherence \cite{38}.  In Fig. \ref{a2}(a), we can observe that for a range of frequency mismatch $0< \Delta \omega < 0.5 $, the system of oscillators attains global synchronization through local synchronization.  Further increasing the frequency mismatch beyond $\Delta \omega >0.5 $, the system attains global synchronization, through local synchronization, local cluster and global cluster states.  From Fig. \ref{a2}(b), we can find that for a sufficiently small value of frequency mismatch ($\Delta \omega \leq 0.34$), global synchronized state is mediated through local cluster states as well as through the global cluster states from individual synchronized states. Here, one cannot observe the existence of chimera states.  Interestingly, we can observe the onset of chimera states only in the range $0.35< \Delta \omega <1.57$.  Beyond this range of frequency mismatch one cannot observe the presence of chimera states.  Thus we conclude that for only a sufficiently large value of frequency mismatch ($\Delta \omega$) one can induce the onset of chimera-like states.  We can also observe that the chimera-like states inherit from out of local cluster states.  Also the chimera-like states mediate between the local cluster states and global cluster states and are quite different from the results in ref. \cite{r4}, where the authors have demonstrated the emergence of chimera states in a system of globally coupled Stuart-Landau oscillators.  Such chimera states have inherited properties from the cluster states in which they originate.  In Figs. \ref{a2}(a) and (b), the locally synchronized region (LS) and local cluster (LC) regions are always stable.  The region-I is the multistability region between the local synchronization and local cluster states.
 In this region, for initial conditions near the globally synchronized state one is lead to complete synchronization among the oscillators.  Also one can check numerically that the chimera-like states and globally clustered states are always stable.
\par  The above discussed results are observed for the case where both the populations are of equal size ($M_1=M_2$).  There arise a question about the robustness of the dynamical states for the case when $M_1\neq M_2$.  To illustrate this, we decrease the number of oscillators ($M_1$) in population-I so that the number of oscillators ($M_2$) in population-II becomes $M_2=N-M_1$.  Then we find the critical value of $n$ ((that is $n=\frac{M_1}{M_2}$)) at which there occurs no qualitative change in the dynamical states as a function of nonisochronicity parameter $c$.  Note that if the value of $n$ is unity it represents that both the populations have equal number of oscillators.  Fig. \ref{log} (a) shows the log-log plot for the critical value of $n$ against the nonisochronicity parameter $c$ for fixed $\varepsilon$ value ( we have fixed $\varepsilon=0.7$ for local synchronization, $\varepsilon=0.9$ for local cluster states, $\varepsilon=1.5$ for chimera-like states, and $\varepsilon=2.3$ for global cluster states).  We can observe that the critical value of $n$ follows the power law relation $n=p c^q$ with c for the local synchronization region and the best curve fit is obtained for the values $p= 0.23108$ and $q= 0.47019$ which is shown by the solid line with ($\blacktriangle$) in Fig. \ref{log} (a).  Similarly the dynamical regions corresponding to local cluster, chimera-like states and global cluster states are illustrated by the solid lines with ($\blacklozenge$), ($\bullet$) and ($\blacksquare$), respectively.  The values of $n$ follow the power law relation with $c$ for the parameter values $p= 0.684708$, $q= 0.140745$ in the local cluster region, $p= 0.0836$, $q= 0.8498$ in the chimera-like states region and $p= 0.3381$, $q= 0.3717$ in the global cluster region.  Similarly, Fig. \ref{log} (b) is plotted for the critical value of $n$ against the coupling strength $\bar{\varepsilon}$ for a fixed $c=3.0$.  Note that $\bar{\varepsilon}=\varepsilon-\varepsilon_e$,  where $\varepsilon_e$ is the earliest coupling strength which is different for different dynamical regions for a fixed nonisochronicity parameter.  To obtain the best curve fit, we subtract the consecutive $\varepsilon$ values with the $\varepsilon_e$ value (that is $\varepsilon_e=0.05$ for local synchronization, $\varepsilon_e=0.8$ for local cluster state, $\varepsilon_e=1.4$ for chimera-like state and $\varepsilon_e=2.25$ for global cluster state).  Here also we can observe that $n$ follows the power law relation with coupling strength $\varepsilon$ for the parameter values $p= 0.916003$, $q= 0.411106$ in the local synchronization region, $p= 0.952379$, $q= 0.060111$ in the local cluster region, $p= 1.00183$, $q=0.599414$ in the chimera-like states region and $p= 1.08815$, $q= 0.433495$ in the global cluster region.
\section{nonlocal interaction}
\par In the previous section, we have analyzed the existence of chimera-like states which represent asynchronous oscillations in one population and synchronous oscillations in the second population in globally coupled oscillators under the influence of frequency mismatch.  To investigate how the dynamics of nonlocally coupled oscillators is affected by the frequency difference between the two populations, we consider the dynamical equations of the following form:
\begin{eqnarray}
\dot{z_j}^{(1,2)}=(1+i\omega^{(1,2)})z_{j}^{(1,2)}-(1- ic)|z_{j}^{(1,2)}|^2 z_{j}^{(1,2)}\nonumber\\+\frac{\varepsilon_1}{M} \sum_{k=1}^{M}(z_k^{(1,2)}-z_{j}^{(1,2)})+\frac{\varepsilon_2}{2P} \sum_{k=j-P}^{j+P} (z_k^{(1,2)}-z_{j}^{(2,1)}),
\label{two}
\end{eqnarray}
where $P$ is the total number of nearest neighbors,  $M$ is the number of oscillators in each population ($M_1=M_2=M$), $N$ is the total number of oscillators in both the populations ($M_1+M_2=N$),  $r=\frac{P}{N}$ is the coupling range.  When $1 < P < N/2$, one has nonlocal coupling.  In the present study, we are interested to analyze the chimera states in two distinct groups of identical populations of nonlocally coupled oscillators.  Here the intra-populations of oscillators are globally connected and the oscillators are nonlocally connected with inter-populations.  For simplicity, we choose the coupling interaction $\varepsilon_1$ = $\varepsilon_2$ = $\varepsilon$ in Eq. (\ref{two}) throughout this section.

\par
\begin{figure}[ht!]
\begin{center}
\includegraphics[width=0.8\linewidth]{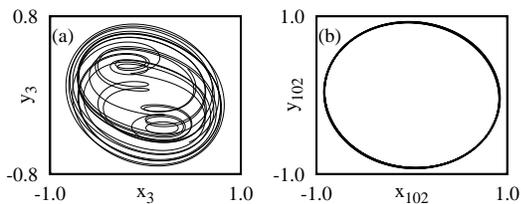}
\end{center}
\caption{Phase portraits of the representative oscillators in imperfectly synchronized state: (a) Represents the non-phase coherent oscillation of the solitary oscillator ($z_3$) in population-I.  (b) Represents the periodic oscillation of the oscillator ($z_{102}$) in population-II.}
\label{n2}
\end{figure}

 In order to analyze the dynamics of the system under nonlocal coupling, we present the spatiotemporal evolution, snapshots and the corresponding frequency profiles $f_j^{(1,2)}$ of the oscillators for various values of coupling strength $\varepsilon$ with fixed coupling range $r=0.4$, $\Delta\omega=0.5$ and nonisochronicity parameter $c=3.0$ in Fig. \ref{n1}.  For a small value of coupling strength $\varepsilon=0.2$, we can observe the local synchronization (Figs. \ref{n1}(a) and (e)) where the oscillators in each of the populations are entrained to two different common frequencies as shown in Fig. \ref{n1}(i).  Note that there is a finite frequency difference between them.  Further increasing coupling strength to $\varepsilon=0.35$, in Figs. \ref{n1}(b) and (f) we can observe the solitary states in population-I where such states represent the fact that some of the oscillators escape from the synchronized group and exhibit non-phase coherent oscillations while we find synchronization in population-II where the oscillators exhibit periodic oscillations.   Figure \ref{n2}(a) shows the phase portrait of the desynchronized oscillator $(z_3)$ which shows the non-phase coherent oscillations of this oscillator and in Fig. \ref{n2}(b) we can observe the periodic oscillation of the synchronized oscillator $z_{102}$.      Another point is that the average frequency of the synchronized oscillators from both the populations are same while the average frequency of the solitary oscillators are different as shown in Fig. \ref{n1}(j).  By increasing the coupling strength to $\varepsilon= 0.5$, the system of oscillators gets split into two groups of synchronized oscillations with periodic motion which is shown in Figs. \ref{n1}(c) and (g).  These two groups are oscillating with common average frequency (as in Fig. \ref{n1}(k)) but with different amplitudes.  These states are designated as cluster states.  Upon increasing the coupling strength to $\varepsilon=1.1$, all the oscillators are entrained to a common frequency with the same amplitude while they are oscillating periodically.  Their spatiotemporal plot, snapshots and frequency profile are shown in Figs. \ref{n1}(d), (h) and (l), respectively.
\begin{figure}[ht!]
\begin{center}
\includegraphics[width=0.9\linewidth]{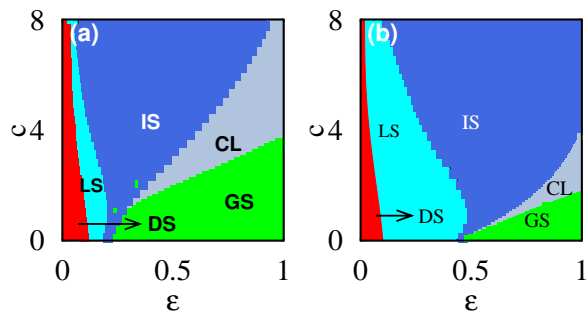}
\end{center}
\caption{(Color online) Two phase diagrams in the ($\varepsilon$,$c$) parametric space (a) for $\Delta \omega=0.5$ and (b) for  $ \Delta \omega=1.5$ with $r=0.4$ for the two population nonlocally coupled Stuart-Landau oscillators. DS represents the desynchronized region, LS represents the local synchronization region, IS shows the imperfectly synchronized states, CL is for cluster state region and GS represents the globally synchronized region.}
\label{nlocala}
\end{figure}
\par In order to explore the results in more detail, we present the global picture in terms of a two phase diagrams in the ($\varepsilon$, $c$) parametric space for two different $ \Delta \omega$ values with fixed coupling range $r=0.4$.  Various dynamical states observed in Fig. \ref{nlocala} are identified by making use of the concept of strength of incoherence \cite{38} (see appendix for more details). From Fig. \ref{nlocala}(a) (for  $ \Delta \omega=0.5$), we can find that the oscillators are initially for low range of $\varepsilon$ (and for all vales of c) exhibiting desynchronized oscillations.  By increasing  the $\varepsilon$ value, the system of oscillators attains local synchronization for a range of $\varepsilon$ values for a given nonisochronicity parameter value.  Upon further increasing the coupling interaction, we can observe the solitary states in population-I and synchronized oscillations in population-II.  This region is marked as imperfectly synchronized state ($IS$) in Fig. \ref{nlocala}.  Note that the IS state is different from the imperfect chimera state reported in ref. \cite{r5} and this state is characterized by a certain small number of solitary oscillators which escapes from the synchronized part of chimera state (where solitary oscillator represents a single repulsive oscillator splitting up from the fully synchronized group). Such escaped oscillators oscillate with different average frequencies.  The IS state is also different from the imperfect chimera state reported in ref. \cite{r6} where the chimeras are characterized by the coexistence of two coherent oscillators and one incoherent oscillator (i.e. rotating with another frequency).  If the chimera state loses its `perfection' in the sense that phases of the synchronized oscillators become not equal, they still rotate with the same average frequency.  In this region, we can also observe the existence of chimera states in population-I while synchronized oscillations in population-II for the specific choice of initial conditions.  If we perturb the system from this specific choice of initial state of the oscillators, the system enters into the imperfectly synchronized states.  This kind of chimera state is different from the state observed in global coupling.  In the case of globally coupled system, one can observe desynchronized oscillations in population-I and synchronized oscillations in population-II, whereas in the present case of nonlocally coupled system, we find the coexistence of coherent and incoherent domains in population-I and coherent oscillations in population-II.  By increasing the strength of the coupling interaction beyond the $IS$ region, the system of oscillators enters into the cluster states (CL).  Further strengthening of the interaction strength leads to global synchronization.  In the CL region, the distribution of initial states near the synchronized solution leads to complete synchronization.  The cluster states are observed for the initial conditions away from the synchronized solution.  Similarly for the case of $\Delta \omega=1.5$ illustrated in Fig. \ref{nlocala}(b), the system of oscillators follows similar transition routes as that of the case  $ \Delta \omega=0.5$ except for the fact that the dynamical regions LS and IS are widened in the $\varepsilon$ space. 

\begin{figure}[ht!]
\begin{center}
\includegraphics[width=1.0\linewidth]{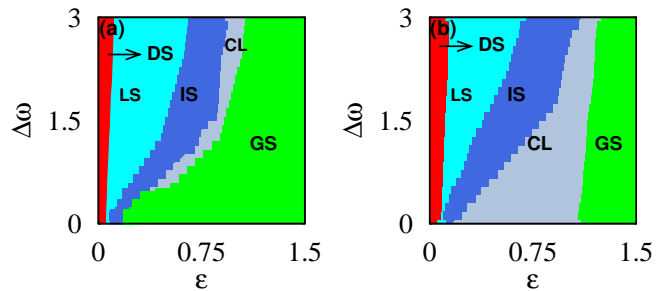}
\end{center}
\caption{(Color online) Two phase diagrams in the ($\varepsilon$,$\Delta \omega$) parametric space (a) for $c=1.5$ and (b) for $c=3.0$ with $r=0.4$ for a system of two distinct identical groups of nonlocally coupled Stuart-Landau oscillators. DS represents the desynchronized region, LS represents the local synchronization, IS shows the imperfectly synchronized states, CL is for cluster state region and GS represents the globally synchronized region.}
\label{nlo}
\end{figure}
\par
\begin{figure}[ht!]
\begin{center}
\includegraphics[width=0.9\linewidth]{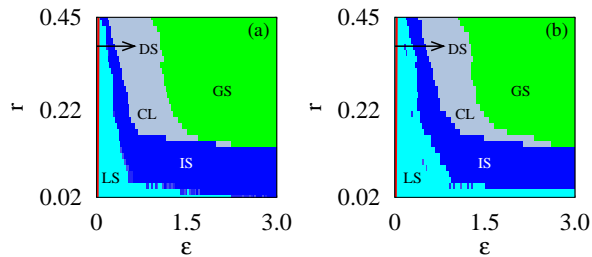}
\end{center}
\caption{(Color online) Two phase diagrams in the ($\varepsilon$,$r$) parametric space (a) for $\Delta \omega=0.5$ (b) for $\Delta \omega=1.5$ with $c=3.0$ for the system (\ref{two}). DS represents the desynchronized region, LS represents the local synchronization, IS shows the imperfectly synchronized states, CL is for cluster state region and GS represents the globally synchronized region.}
\label{re}
\end{figure}

 Next we analyze whether the emergence of imperfectly synchronized states (or chimera states for spatially prepared initial conditions) are robust to an increase of frequency mismatch between the two populations.  To illustrate the results, we plotted the two parameter phase diagrams in the ($\varepsilon, ~\Delta \omega$) parametric space in Figs. \ref{nlo}(a,b) for $c=1.5$ and $c=3.0$ values, respectively, with fixed coupling range $r=0.4$.  In Fig. \ref{nlo}(a), initially the system of oscillators is desynchronized for small values of coupling strength.  By increasing $\varepsilon$ in the range $0<\Delta\omega<0.5$, the system attains global synchronization (GS) through the locally synchronized states (LS) and imperfectly synchronized (IS) states.  For $\Delta\omega \ge 0.5$, the system attains global synchronization (GS) through the cluster states (CL) in addition to the locally synchronized states (LS) and imperfectly synchronized states (IS).  By increasing the value of $c$ to $c=3.0$, the system attains global synchronization (GS) through the locally synchronized states (LS), imperfectly synchronized states (IS) and cluster states (CL) for all values of frequency mismatch between the populations as shown in Fig. \ref{nlo}(b).  In Figs. \ref{nlo}(a) and (b), the regions IS and CL are multistable regions for the reason that even for small perturbations of initial states from the synchronized state leads to the existence of imperfectly synchronized states and cluster states in the regions IS and CL, respectively.  Also in the IS region, we can observe the onset of chimera states for specific choice of initial conditions.  We can observe that the existence of chimera state is independent of frequency mismatch between the two populations under nonlocal coupling, whereas in the case of global coupled system the existence of chimera state is obtained for only a range of frequency mismatch.  We have also analyzed the results for a wide range of nonlocal coupling by plotting the two parameter phase diagram in the ($\varepsilon, ~r$) parametric space with two different $\Delta \omega$ values that is $\Delta \omega=0.5$ and $\Delta \omega=1.5$ for fixed $c=3.0$.  Figures \ref{re}(a) and (b) clearly illustrate the existence of imperfectly synchronized states (or chimera states for spatially prepared initial conditions) and other dynamical states for a wide range of nonlocal coupling independent of the frequency mismatch. 
\section{Conclusion}
In summary, we have illustrated the existence of different dynamical states in globally and nonlocally coupled systems of two distinct groups of identical populations of oscillators.  We have analyzed the impact of frequency mismatch between the two populations.  An interesting feature here is that only for certain ranges of frequency mismatch, one can observe the onset of chimera-like states in globally coupled network and the oscillators are coupled with equal strength coupling interaction.  These states are different from the chimera states observed in coupled identical populations where the total number of oscillators are split into coexisting domains of synchronized and desynchronized oscillations.  We have found the results are robust for certain range of unequal size populations (both the populations contain unequal number of oscillators).  In addition, we have found the existence of multistable regions depending on the initial state of the oscillators.   
\par Further we have extended our study to analyze the dynamics of nonlocally coupled oscillators where we find the existence of chimera state for spatially prepared initial conditions.  In contrast to the global coupling, existence of chimera states is independent of the increase of frequency mismatch between the two populations.  In contrast to the case of global coupling, multistability does not exist in nonlocally coupled system except in the cluster state region.  Finally, it may be an interesting open problem to extend this study to a network of more than two populations, which we are pursuing currently.
\section*{Acknowledgements}
The work of VKC is supported by the SERB-DST Fast Track scheme for young scientists under Grant No.YSS/2014/000175.  The work of MS forms part of a research project sponsored by DST, Government of India.  ML acknowledges the financial support under a NASI  Platinum Jubilee Senior Scientist Fellowship program.
\appendix
\section{CHARACTERISTIC MEASURE FOR STRENGTH OF INCOHERENCE}
\par To identify the nature of different dynamical states, we look at the strength of incoherence of the system a notion introduced recently by Gopal, Venkatesan and two of the present authors\cite{38}, that will help us to detect interesting collective dynamical states such as synchronized state, desynchronized state, and the chimera state.  For this purpose we introduce a transformation $w_j^{(1,2)}=x_j^{(1,2)}-x_{j+1}^{(1,2)}$ \cite{38}, where $j=1,2,3,...,N$.  We divide the oscillators into $K$ bins of equal length $l=N/K$ and the local standard deviation $\sigma_l(m)^{(1,2)}$ is defined as   
\begin{eqnarray} 
\sigma_l(m)^{(1,2)}=\langle(\overline{ \frac{1}{l}\sum_{j=l(m-1)+1}^{ml} \vert w_j^{(1,2)}-\overline{w^{(1,2)}}\vert^2})^{1/2}\rangle_t,\nonumber\\
 m=1,2,...K.
\label{sig}
\end{eqnarray}
\par From this we can find the local standard deviation for every $K$ bins of oscillators that helps to find the strength of incoherence \cite{38} through
\begin{equation} 
S^{(1,2)}=1-\frac{\sum_{m=1}^{K} s_m^{(1,2)}}{K},s_m^{(1,2)}=\Theta(\delta- \sigma_l(m)^{(1,2)}),
\label{soi}
\end{equation}
where $\delta$ is the threshold value which is small. Here $\langle ... \rangle_t$ represents the average over time.  When $\sigma_l(m)^{(1,2)}$ is less than $\delta$, $s_m^{(1,2)}$ takes the value $1$, otherwise it is $0$.  Thus the strength of incoherence measures the amount of spatial incoherence present in the system which is zero for the spatially coherent  synchronized state.  It has the maximum value, that is $S^{(1,2)}=1$, for the completely incoherent desynchronized state and has intermediate values between 0 and 1 for chimera states and cluster states. 

\end{document}